\documentclass{elsart3-1}

\usepackage{graphicx}
\usepackage{float}

\usepackage{amssymb}
\usepackage{amsmath}
\usepackage[english]{babel}
\usepackage{color}
\usepackage[justification=centering]{caption}

\newtheorem{e-proposition}[theorem]{Proposition}

\newtheorem{e-definition}[theorem]{Definition\rm}

\renewcommand{\theequation}{\arabic{equation}}
\def\og{\leavevmode\raise.3ex\hbox{$\scriptscriptstyle\langle\!\langle$~}}
\def\fg{\leavevmode\raise.3ex\hbox{~$\!\scriptscriptstyle\,\rangle\!\rangle$}}

\begin{document}

\begin{frontmatter}

\selectlanguage{english}

\title{A Lattice-Boltzmann method for the interaction between
mechanical waves and solid mobile bodies.}
\selectlanguage{english}
\author[Unal]{A. M. Velasco},
\ead{amvelascos@unal.edu.co}
\author[Unal]{J. D. Mu\~noz}
\ead{jdmunozc@unal.edu.co}

\address[Unal]{Simulation of Physical Systems Group, Department of Physics, Universidad Nacional de Colombia, Cra 30 No. 45-03, Ed. 404, Of. 348, Bogotá D.C., Colombia}

\maketitle
\begin{abstract}
The acoustic waves generated by moving bodies and the movement of bodies by acoustic waves are central phenomena in the operation of musical instruments and in everyday's experiences like the movement of a boat on a lake by the wake generated by a propelled ship. Previous works have successfully simulated the interaction between a moving body and a lattice-Boltzmann fluid by immersed boundary methods.  Hereby, we show how to implement the same coupling in the case of a Lattice-Boltzmann for waves, i.e. a LBGK model that directly recovers the wave equation in a linear medium, without modeling fluids. The coupling is performed by matching the displacement at the medium-solid boundary and via the pressure, which characterizes the forces undergone by the medium and the immersed body. The proposed model simplifies the preceeding immersed boundary methods and reduces the calculations steps. The method is illustrated by simulating the movement of immersed bodies in two dimensions, like the displacement of a two-dimensional disk due to an incoming wave or the wake generated by a moving object in a medium at rest. The proposal consitutes a valuable tool for the study of acoustical waves by lattice-Boltzmann methods.  
\end{abstract}
\end{frontmatter}
\selectlanguage{english}
\section{Introduction}

Lattice-Boltzmann models are useful tools to simulate a wide variety of processes in fluid dynamics, from capillarity and hydrodynamic instabilities to porous media and general rheology. Furthermore, in recent years its applications have been extended to other systems whose behavior is governed by a set of conservation laws, like electrodynamics or waves \cite{MendozaMunozElectro,ChopardDroz,GuangwuWaves}, extending the concept of lattice-Boltzmann from a mesoscopic model to simulate fluids to a general numerical scheme to solve more general differential equations. The main advantage of this scheme is its local nature, because all variables needed to compute the time evolution of a single node before writing them in the neighboring nodes are present in the node itself, making them perfect to run parallel on graphic cards. Furthermore, due to its mesoscopic approach, the inclusion of physical ingredients like inter-particle interactions or body forces can be done naturally. Because of its versatility and parallel nature, lattice-Boltzmann methods has gained the interest of a wide range of research areas and industrial applications.

An important issue with great scientific and industrial interest is the coupling between solid (rigid or flexible) moving objects and a surrounding fluid. However, this coupling is not an easy task if the fluid is modeled by a lattice-Boltzmann, due to its Eulerian nature, in contrast with the Lagrangian character of the moving object. Thus, it is necessary to establish a proper communication between the fluid and the solid, through a consistent set of boundary conditions. In 2014 Favier, Revell and Pinelli \cite{ref1} proposed a Immersed Boundary method for a lattice-Boltzmann model for fluids, inherit from that for Finite Element methods, to simulate the interaction between rigid an flexible objects. With this model, Favier \textit{et. al.} studied the drafting, kissing and tumbling process of two sediment particles, and the dynamics of a flexible filament immersed in a fluid flux. Despite the great usefulness of this method, it requires a complicated algorithm to compute the two directional solid-fluid interaction and, it requires an additional lattice-Boltzmann calculation for every time step, increasing CPU time.
Because Favier's approach is relatively new, it has not been extended to systems apart of fluids, like electrodynamics or mechanical waves. 
Nevertheless, the wave-solid coupling is fundamental to describe systems like the wake generated by a boat on the surface of a lake, the sound generated by a musical instrument or the vibrational response of a solid object receiving the impact of a shock wave.

In this work, we extend the coupling proposed by Favier \textit{et. al.} to reproduce the interaction between a lattice-Boltzmann model for acoustic waves and solid objects. Section \ref{model} reviews the lattice-Boltzmann method for mechanical waves \cite{ChopardDroz} and introduces the new approach to simulate the wave-solid interactions. Section \ref{Results} uses the new model to simulate the movement of a solid disk due to the impact of a mechanical wave and, the wake generated by mobile objects with different shapes immersed in a compressible media. Finally, we summarize the main results and conclusions in Section \ref{Conclusions}.

\section{lattice-Boltzmann method for the simulation of mechanical waves}\label{model}

\begin{figure}[H]
\centering
\includegraphics[width=0.4\textwidth]{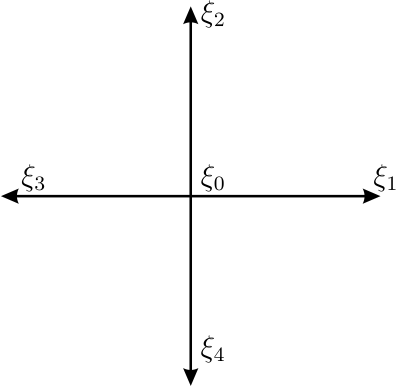}
\caption{D2Q5 (2 dimensions, 5 velocities including zero velocity)(Left) and D2Q9 (Right) discrete velocities schemes.}
\label{Fig.Velocities}
\end{figure}

The lattice-Boltzmann model for acoustic waves in two dimensions divides the space into square cells of equal size, as usual. There is also a discrete set of velocity vectors $\vec{\xi_{i}}$ connecting each cell with its neighbors (Fig. \ref{Fig.Velocities}). Furthermore, each cell at poition $\vec{x}$ contains a set of distribution functions $f_{i}(\vec{x},t)$, each one attached to a discrete velocity and representing the probability density of finding a fluid particle there with such velocity at a given time.  The physical information of the system is obtained by computing the statistical moments of those distribution functions, which correspond to the fields of macroscopic variables. Next, these macroscopic variables are used to compute equilibrium values $f_i^{eq}$ for the distribution functions. The distribuion functions evolve according to the discrete kinetic lattice-Boltzmann equation in the BGK approximation \cite{BGK},
\begin{equation}
f_{i}\left(\vec{x}+\vec{\xi}_{i}\Delta_{t},\,\vec{\xi}_{i},\, t+\Delta_{t}\right)-f_{i}\left(\vec{x},\,\vec{\xi}_{i},\, t\right)=-\frac{\Delta t}{\tau}\left[f_{i}\left(\vec{x},\,\vec{\xi}_{i},\, t\right)-f_{i}^{eq}\left(\vec{x},\,\vec{\xi}_{i},\, t\right)\right]\,.\label{eq:evolution}
\end{equation}
Here, $\tau=0.5$ is known as the {\it relaxation time}, and the index $i$ runs over the set of discrete velocities. Finally, each distribution function travels to the neighboring cell with the velocity vector it is associated to. The procedure repeats for each timestep.

The statistical moments of the distribution functions give the macroscopic fields. In our case, these interesting quantities will be the pressure,
\begin{equation}\label{RHO}
	P=\sum_i f_i^{eq} \quad ,
	\end{equation}
a vector field $\vec J$	
	\begin{equation}\label{J}
	\vec J=\sum_i \vec \xi_i f_i^{eq} \quad , 
\end{equation}
whose physical meaning will be discussed later, and a stress tensor, given by
\begin{equation}\label{PI}
	\Pi^{(0)}=\sum_i  \vec \xi_i \otimes \vec \xi_i f_i^{eq}\quad .
	\end{equation}
As shown in \cite{ChopardDroz}, these fields fulfill the following conservation laws
\begin{equation}\label{1Conservation}
	0=\frac{\partial P}{\partial t}+\nabla\cdot\vec J\quad \quad ,
\end{equation}
\begin{equation}\label{2Conservation}
	0=\frac{\partial\vec J}{\partial t}+\nabla\cdot\vec \Pi^{(0)}\quad.
\end{equation}
Note that if we could define $f_{i}^{eq}$ such that 
\begin{equation}
\Pi^{(0)}=\begin{pmatrix}
c^2 P & 0\\ 
0 & c^2 P
\end{pmatrix}\quad,
\label{Stress}
\end{equation}
with $c$ a constant, we would have 
\begin{equation}\label{2Conservation2}
	\nabla\cdot\Pi^{(0)}=c^2\nabla P\quad,
\end{equation}
and Eq. (\ref{1Conservation}) and (\ref{2Conservation}) could be combined to obtain a wave equation for $P$,
\begin{equation}\label{WaveEquation}
	0=\frac{\partial^2 P}{\partial t^2}-c^2\nabla^2 P\quad,
\end{equation}
The constant $c$ shows to be the wave velocity. In addition, because the pressure in an acoustic wave is related with the mean displacement $D$ of the particles in the fluid through the relation $P=-B\nabla\cdot \vec{D}$ where $B$ is the bulk modulus \cite{Elmore}, Eq. \ref{1Conservation} say us that the macroscopic quantity $\vec{J}$ is related with the displacement velocity  
\begin{equation}\label{DefJ}
	\vec{J}=B\frac{\partial \vec{D}}{\partial t} .
\end{equation}

The problem reduces, therefore, to propose the appropriate equilibrium distribution function to guarantee at the same time the equations \ref{RHO}-\ref{PI} and the condition \ref{Stress}. This can be done if the equilibrium distribution function takes the form (See Appendix A),
\begin{equation}
f^{eq}_{i}=\left\{\begin{matrix}
P+\frac{Pc^2}{c^2_s}\left(w_{0} - 1\right) & \mathrm {if}\,  i= 0\\ 
\frac{w_{i}}{c_s^2}\left(P c^2+ \vec{\xi}_i \cdot \vec{J}\right)  &\mathrm {if} \, i\neq 0
\end{matrix}\right.
 \end{equation}

\section{lattice-Boltzmann Immersed Boundary method for waves}
Our goal now is to couple the previous model for acoustic waves with the dynamics of a rigid body. Part of this approach is based on the immersed boundary method proposed by Favier, Revell and Pinelli \cite{ref1} for the coupling between immersed solid structures and fluids. We describe our model below.

\subsection{Action of the wave on the solid.}
In a lattice-Boltzmann method the information is located on the lattice nodes, which remain fixed in space during the simulation. In contrast, the solid immersed object moves in space as time goes on. Let us divide the boundary of the solid immersed object into a set of $M$ discrete surfaces, each one almost flat, with surface vector $\Delta\vec{s}_k$ pointing out of the object and centered at point $\bar{\vec{R}_k}$. Let us remember that we know the wave presure on each lattice node, 
\begin{equation}
P=\sum_i f_i^{eq}\quad;
\end{equation}
therefore, it is only necessary to determine how this pressure acts on the body surface. For this end, we interpolate among the nodes to find the value of the pressure at each point $\bar{\vec{R}_k}$ with a kernel function $\tilde{\delta}(r)$, 
\begin{equation}
\label{Pinterpmidpoint}
P\left(\bar{\vec{R}_k}\right)\simeq\sum_l P\left(\vec{y}_l\right)\tilde{\delta}\left(\vec{y}_l-\bar{\vec{R}_k}\right)\quad,
\end{equation}
where the sum extends on all positions $y_l$ of the lattice Boltzmann nodes. In the present work we will use the kernel  $\tilde{\delta}$ defined as (Fig. \ref{Fig.Scheme})

\begin{equation}
\tilde{\delta}\left(r\right)=\left\{\begin{matrix}
\frac{1}{6}\left(5-3\left|r\right|-\sqrt{-3\left(1-\left|r \right| \right)^{2}+1}\right) & \mathrm {if}\,  1/2\leq \left|r\right| \leq 3/2\\
\frac{1}{3}\left(1+\sqrt{-3r^2+1}\right) &\mathrm {if} \, \left|r\right| \leq 1/2\\
0 & \mathrm{otherwise}
\end{matrix}\right.
\label{kernel}
 \end{equation}

The total force exerted by the fluid on the immersed solid is, therefore,
\begin{equation}
\vec{F_{ws}}=-\sum^{M}_{k=1}P\left(\bar{\vec{R}_k}\right)\Delta\vec{s}_k  \quad. 
\end{equation}
Once the force acting on the object has been found, we can use any integrator to move the object and update its velocity and position on each timestep.

\subsection{Action of the solid on the wave}
\begin{figure}[h!]
\centering
\includegraphics[width=0.6\textwidth]{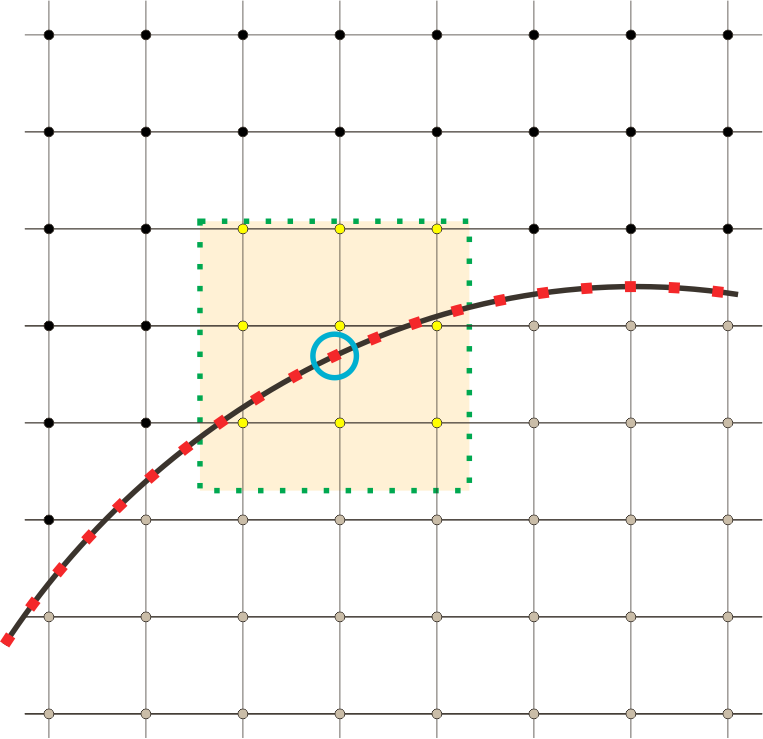}
\caption{Influence area for a given discrete boundary point (within the circle)}
\label{Fig.Scheme}
\end{figure}

Following the proposal of Favier, Revell and Pinelli, we introduce a forcing term  at the discrete boundary solid points which ensures that the displacement velocity of the wave in the boundary matches the velocity of the discrete solid. Since the first moment $\vec{J}$ of the lattice-Boltzmann model is proportional to the velocity displacement $\vec{D}$ (Eq. \ref{DefJ}), the forcing term will be indeed a change of the quantity $\vec{J}$, i. e.  
\begin{equation}\label{Force}
\vec{F}(\vec{R}_k)=\frac{B\vec{U}_k-\mathcal{I}\left(\vec{J}\right)(\vec{R}_k)}{\Delta t}\quad.
\end{equation}
Here $\vec{R}_k$ is the position of the $k$-th boundary solid point, $\vec{U}_k$ is the velocity of that solid point (obtained from integrating its movement equations), and $\mathcal{I}\left(\vec{J}\right)(\vec{R}_k)$ is the interpolated value of $\vec{J}$ at $\vec{R}_k$,
\begin{equation}
\label{Jinterpdisc}
\mathcal{I}\left(\vec{J}\right)(\vec{R}_k)=\sum_l \vec{J}\left(\vec{y}_l\right)\tilde{\delta}\left(\vec{y}_l-\vec{R}_k\right)\quad,
\end{equation}
where $\vec{y}_l$ is the position of the $l$-th lattice node within the influence area (which, for the kernel Eq.(\ref{kernel}) is just three cells width, see Fig. \ref{Fig.Scheme}).

Because this forcing term is located at the discrete boundary points, it is necessary to spread it out to the lattice nodes in order to force the wave simulated by the lattice-Boltzmann. This can be done by using the same interpolation kernel. The force on the $l$-th lattice node is
\begin{equation}
\label{Forcelattice}
\vec{f}\left(\vec{y}_l\right)=\sum_{k=1}^M \vec{F}_{bp}(\vec{R}_k)\tilde{\delta}\left(\vec{y}_l-\vec{R}_k\right)\epsilon_k\quad,
\end{equation}
where $\epsilon_k$ is a value that ensures the consistency between the interpolation and posterior spreading process, as follows. Let us fix a force at a certain solid point $\vec{R}_k$. Spread the force forth to the lattice nodes Eq. \eqref{Forcelattice} and back to the point $\vec{R}_k$ Eq. \eqref{Jinterpdisc}. If the returned value is the same as the original one, the value of $\epsilon_k$ is right. 

Unlike the model proposed by Favier, Revell and Pinelli, the force Eq. \eqref{Forcelattice} will not be included in the collision term, because $\tau=0.5$, but just in the definition of the macroscopic quantity $\vec{J}$,
\begin{equation}
\label{Jshift}
\vec{J}\left(\vec{y}_l\right)=\sum_i\vec{\xi}_if^{eq}_i+\frac{\Delta t}{2}\vec{f}\left(\vec{y}_l\right)
\end{equation} 
That completes the wave-solid interaction.

The proposed method is completely general and can implement the interactions between wave and solid in many different phenomena. Furthermore, if desired, it also allows to propagate waves inside the solid, with the same or a different propagation speed, as we will illustrate with some examples. 



\section{Simulations and discussion}\label{Results}

\begin{figure}[H]
\centering
\includegraphics[width=0.6\textwidth]{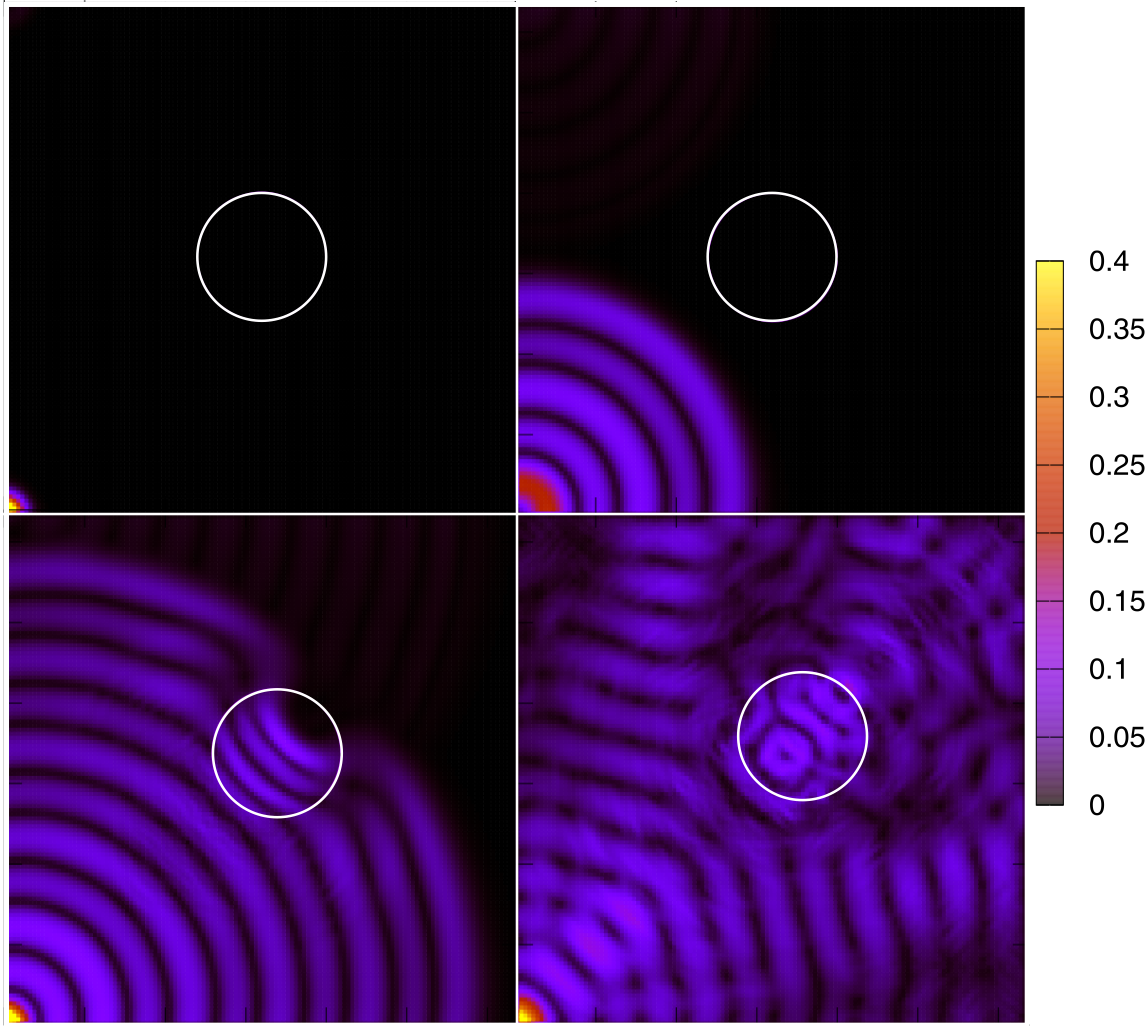}
\caption{Displacement of a solid disc generated by a polar wave}
\label{Fig.Source}
\end{figure}

In order to prove our new model, we first study the effect of a  mechanical wave on the movement of an object initially at rest, to this aim,  we place a two-dimensional solid disk at the center of the computational domain and we define a sinusoidal perturbation coming from a corner of the domain. The boundary conditions at the walls are bounce-back with a damping factor equal to $0.6$ in order to control the amplitude of the perturbation and achieve an stationary state. The resulting movement and the pressure profile is shown in the Figure \ref{Fig.Source}

\begin{figure}[H]
\centering
\includegraphics[width=1\textwidth]{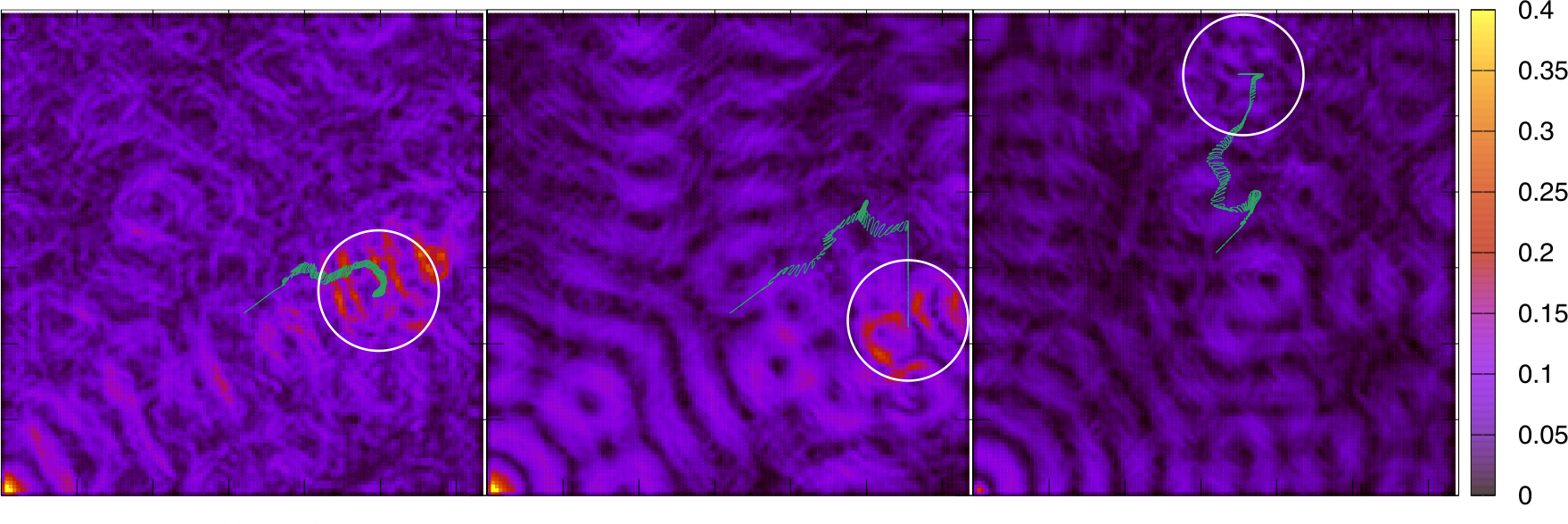}
\caption{Trajectory of the disk after 3000 time steps for frequencies $5.25$, $5.25$ and $5.75$, and velocity ratios $0.5$, $0.5$, and $0.6$ respectively from left to right}
\label{Fig.Path}
\end{figure}

Note how the disc is displaced from the center in the same direction of the wave and the circular wavefront is disturbed by the change of velocity within the solid (bottom-left image). In order to track the displacement and make a more detailed study of the interaction, we trace the trajectory of the disk for simulations of different wave frequency and velocities ratio $n=v_{s}/c$ where $v_{s}$ is the wave velocity inside the solid and $c=0.1$ in the medium, Figure \ref{Fig.Path} shows the result

\begin{figure}[H]
\centering
\includegraphics[width=0.6\textwidth]{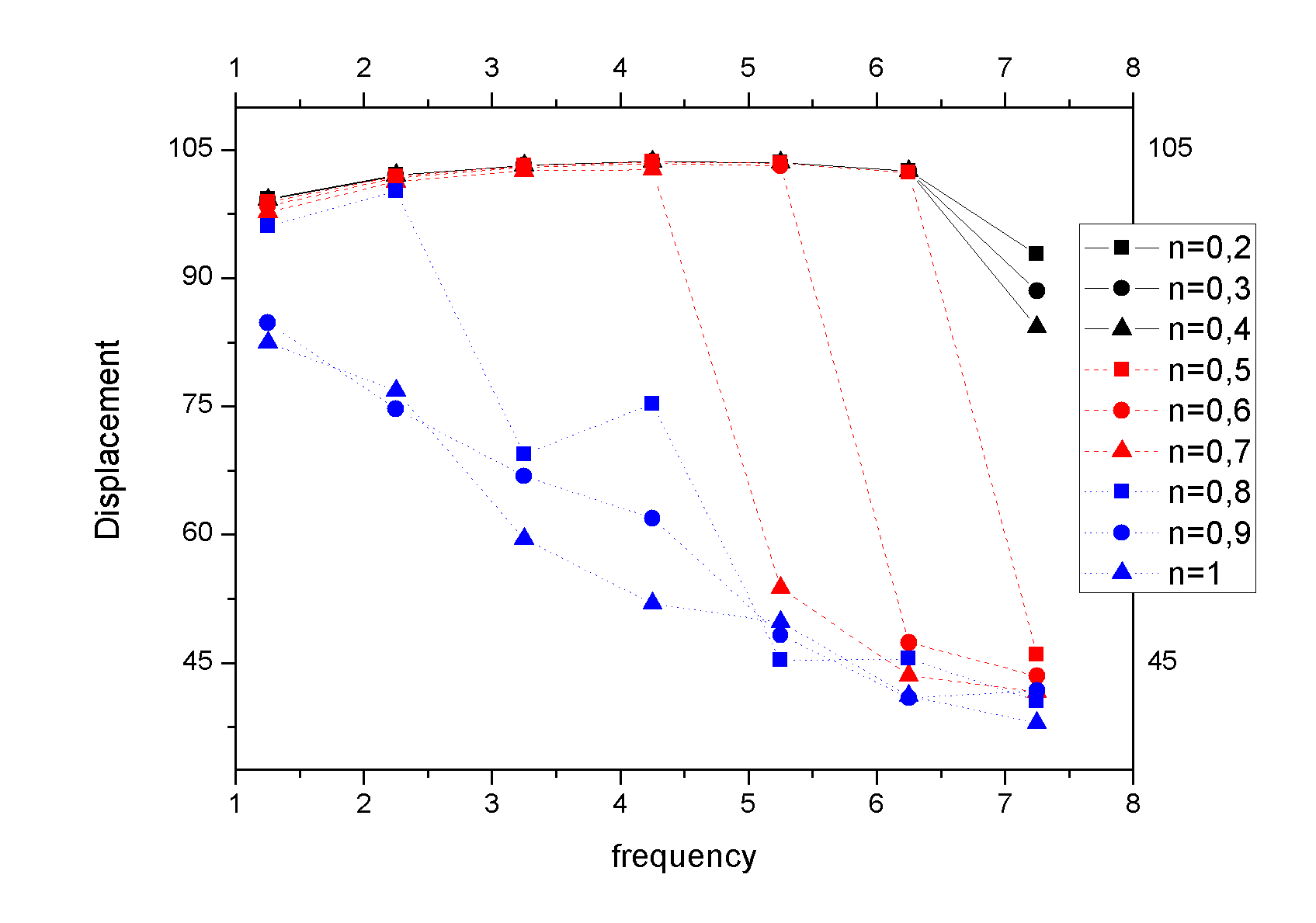}
\caption{Displacement of the disk as a function of the wave frequency for different values of $n$}
\label{Fig.Relation}
\end{figure}

Where the frequency and velocity is measured in units of time steps $\Delta t = 1$ computing clock pulse and unitary cells. Note that the trajectory greatly depends on the value of $n$ and the frequency of the wave. The Figure \ref{Fig.Relation} shows the relation between the displacement of the disk as a function of the frequency, for different values of $n$ of a straight line trajectory.

As expected, disks whose wave velocity at the interior is low show a greater displacement than disks that do not present such a high resistance to the pressure vibrations. Respect to the frequency the displacement has a decreasing behavior, higher frequencies displaces the disk a shorter distance. This effect is stronger when the velocity of the wave inside the disk is 0.8, 0.9 and 1 times $c$ (blue dotted lines), i.e. when the material has a low rigidity, for values of $n$ equal to 0.5 0.6 and 0.7,(red dashed lines) there is a cutoff frequency from which the wave crosses the disk displacing it a considerably shorter distance. Note that the lower the velocities ratio is, the higher the cutoff frequency gets, this means that it is necessary a higher frequency to overpass a more rigid object. Furthermore, the relation between the displacement and the frequency for values of $n<0.7$ is increasing for frequencies under the cutoff, even for $n$ equal to 0.2, 0.3 and 0.4, where the skip is not too abrupt and the cutoff frequency does not appear clearly.\\

\begin{figure}[H]
\centering
\includegraphics[width=1\textwidth]{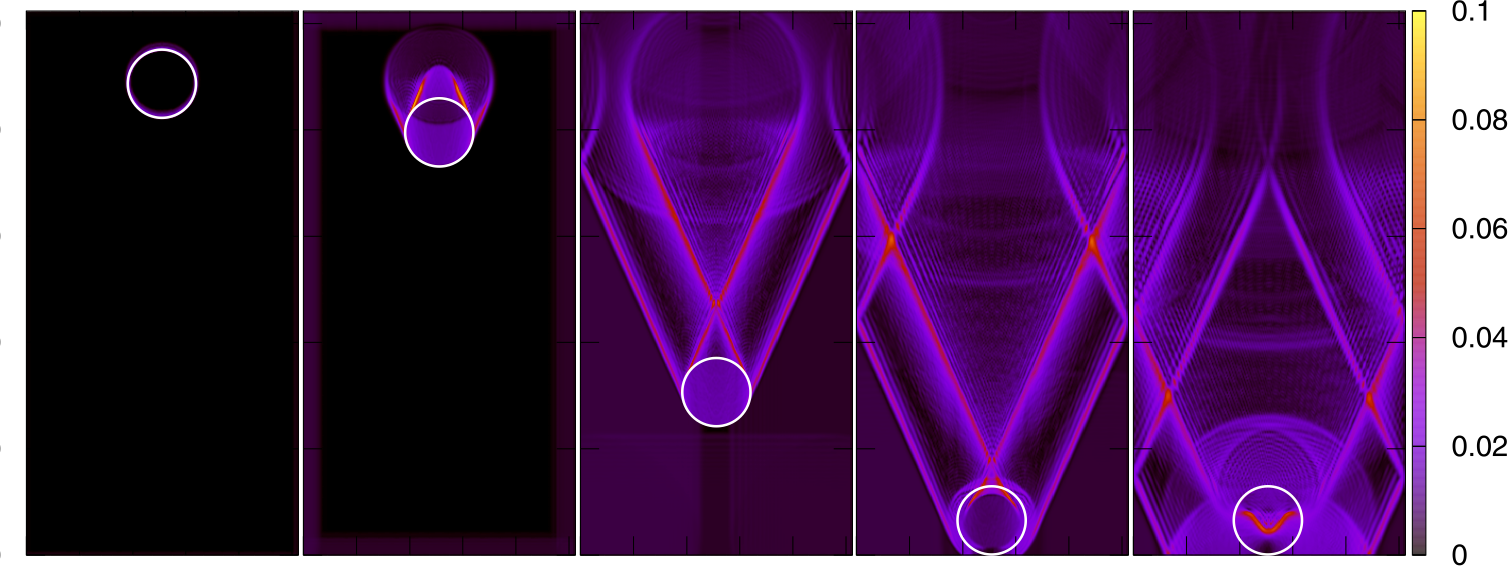}
\caption{Accelerated disk moving downwards in a medium. Frames taken at 10, 100, 600, 900, and 1200 time steps}
\label{Fig.Wake}
\end{figure}

As a second part of our proves, now we start from a medium at rest and apply a force to the solid object in order to study the generated wakes. Figure \ref{Fig.Wake}  shows the wake generated by a disk with $n=0.6$ at different time steps.

\begin{figure}[H]
\centering
\includegraphics[width=0.6\textwidth]{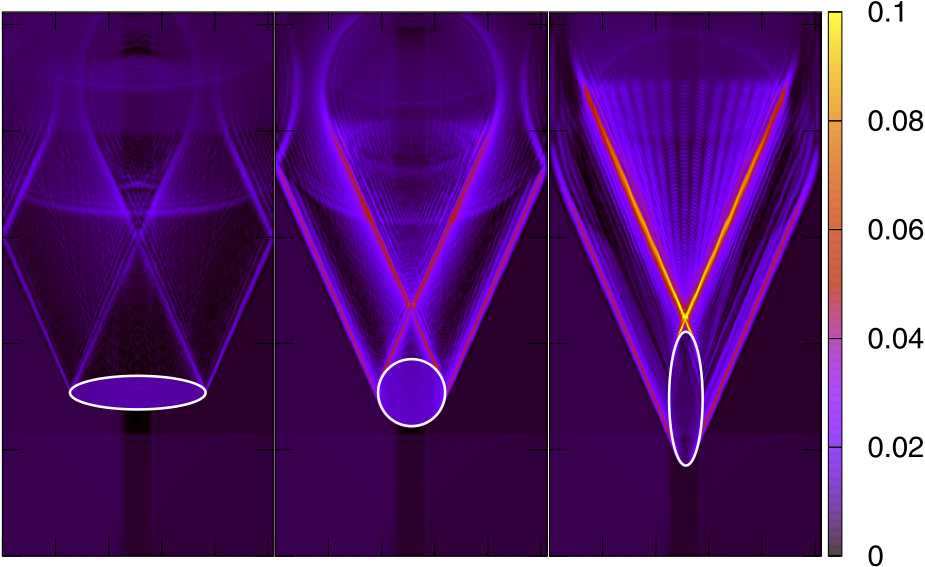}
\caption{Wake generated from accelerated objects of different shapes.}
\label{Fig.Shapes}
\end{figure}

The resulting wake is qualitatively as expected, it expands as the disk advances and a new wave moving out of the disk is generated when it hits the bottom wall. It is also interesting to study how the shape of the solid object influences the creation of the wake. Simulations of the same system for different shapes of the solid are shown here in Figure \ref{Fig.Shapes}.

The results show that a sharper front of the moving object generates a stronger wake, i. e. the vertical ellipse generates a higher pressure behind it and the triangle formed by the intersection of the both sides of the wake tends to disappear when the solid becomes sharper. 

\begin{figure}[H]
\centering
\includegraphics[width=0.38\textwidth]{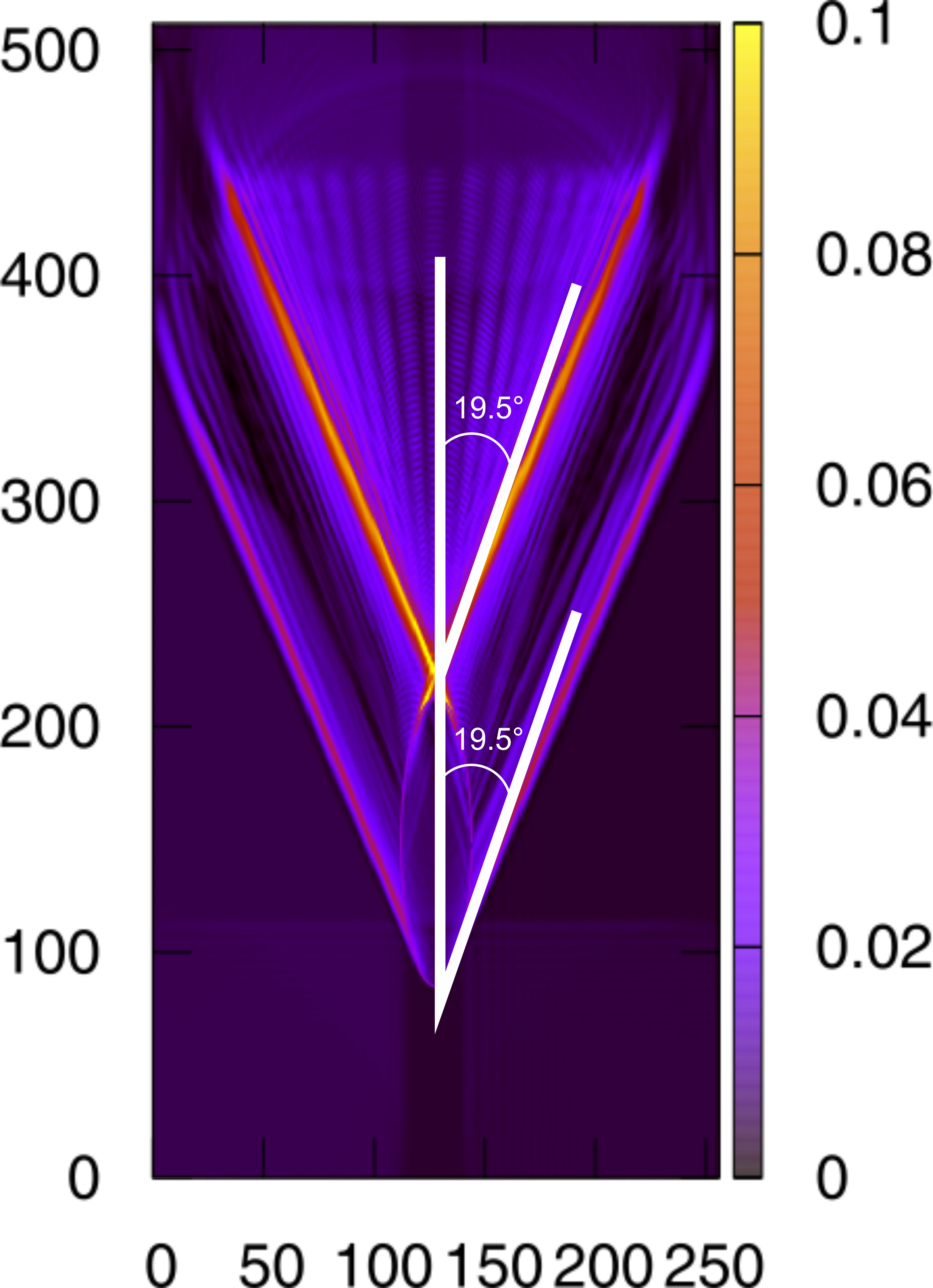}
\caption{Comparison between the obtained wake angle and the expected Kelin's Angle.}
\label{Fig.Kelvin}
\end{figure}

The value of the angle formed by the wake respect with the vertical axis is compared with the theoretical expextations, this angle, known as Kelvin's angle  \cite{Kelvin} should be of 19.5$^{\circ}$ , our results shown a deviation of 15$\%$ approximately, which can be due to the low order aproximation on the lattice velocities discretization. Fig. \ref{Fig.Kelvin}

In addition, we study the convergence of our model by measuring the displacement of a rigid object due to the interaction with a pressure pulse as a function of the cell size $\varepsilon$, the result is shown here

\begin{figure}[H]
\centering
\includegraphics[width=0.6\textwidth]{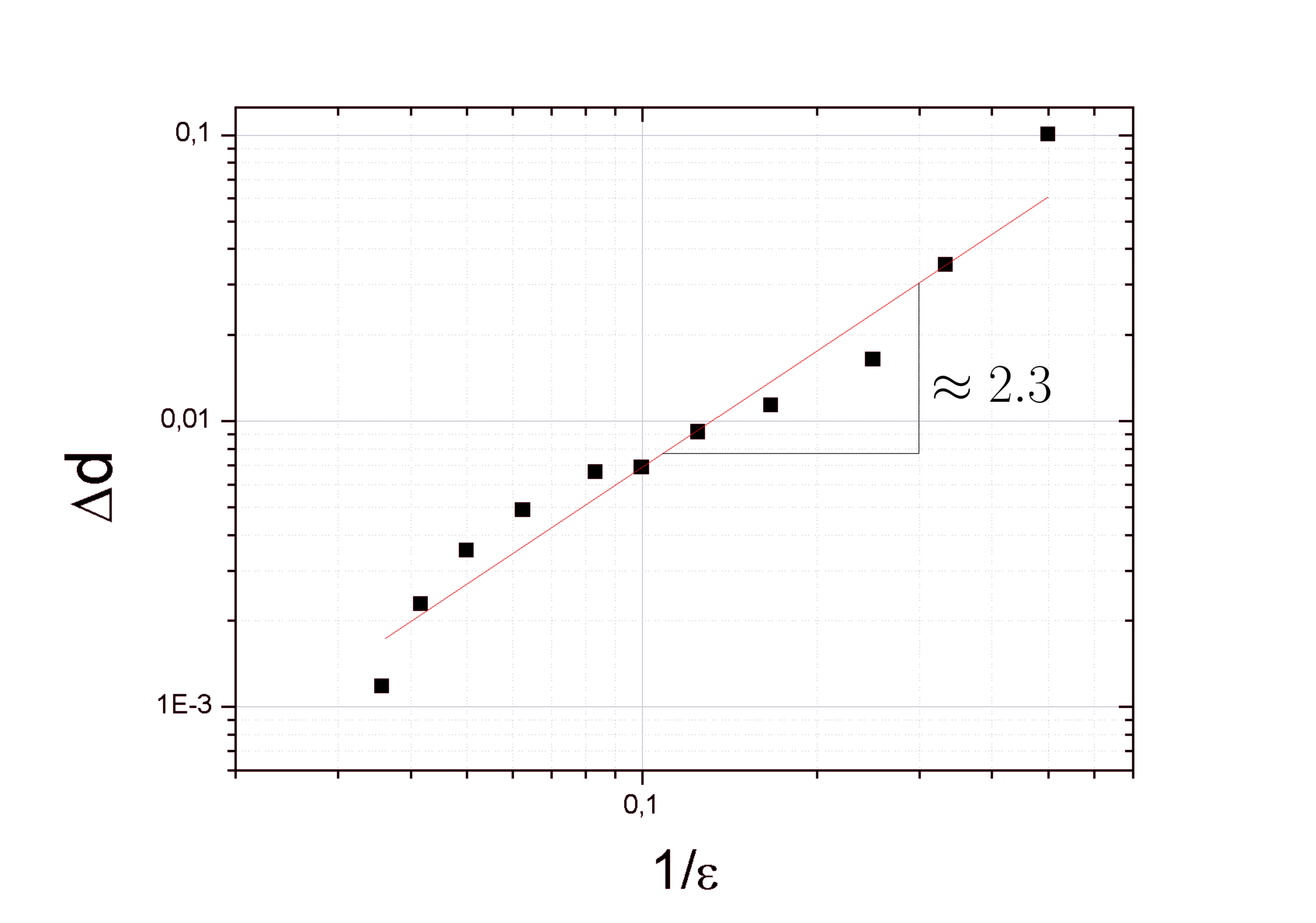}
\caption{Logarithmic graph of the error in displacement of an object as a function of the lattice resolution.}
\label{Fig.Error}
\end{figure} 

where $\Delta d$ is the difference between the meassured displacement and a theoretically known value. Fig. \ref{Fig.Error} shows that our model has second order accuracy, since the slope found was 2.3 approximately.

\section{Conclusions}\label{Conclusions}

A lattice-Boltzmann method (LBM) for the study of the coupled behaviour between solid objects and pressure waves was developed and tested. The proposal implements a LBM for the direct simulation of the wave equation (without any fluid mechanics) and designs a proper forcing term to couple with the moving boundaries of immersed objects, by following a similar strategy to the  immersed boundary method proposed by Favier \textit{et. al.} for fluids []. The effect of the medium on the solid object was introduced by computing the normal pressure on the object boundary exerted by the medium. In turn, the moving object generates pressure waves inside the medium. This effect is modelled by matching the velocity of the solid boundary with that of the medium through a forcing term included in the LBM scheme. In addition, the wave speed inside the solid objects can be chosen as desired. \\

The one dimensional displacement of a solid object on a fixed time as a function of the frequency of an incoming wave was studied for different values of the wave speed inside the solid. Our results show three different behaviours: First, when the wave speed inside the object is similar to the one in the medium, the object's displacement monotonically decreases when the frequency increases. Second, for objects whose internal wave speed lies between 0.5 and 0.7 times the wave speed in the medium, the displacement remains almost constant for frequencies below a cutoff value; above this wave can cross the solid easily without displacing it a considerable amount, this cutoff frequency gets higher when the velocity ratio is lower, i. e. when the solid has a stronger resistance to the vibrations inside. Third, when the wave speed inside the objects lies below 0.5 times the wave speed in the medium, the cutoff frequency is not longer evident, but the displacement still falls for higher frequencies. The bi-dimensional trajectories followed by a disc under the influence of a circular wave in a squared cavity are also presented. \\

The effect of moving objects on the surrounding medium was also studied with our method. The generated wake satisfies the theoretical expectation showing a Kelvin's angle with an error around 15$\%$. This is due probably to using a D2Q5 discrete velocity set for the LBM. The effect of using other velocity sets as D2Q9 would be a nice topic for future research.\\

The model is easy to implement and allows us to control some physical variables like the pressure of the medium and the wave speed inside and outside the solid object. In addition, the shape of the object is controlled by the position of the boundary points, whose dynamics is updated every time step. Due to this fact, the shape of the object can be easily modified in time, and the simulation of flexible objects could be included in a natural way just by changing the dynamics of the rigid body. This proposal constitutes, therefore, a valuable tool to simulate the interaction between pressure waves and immersed objects.


\begin{thebibliography}{9}
\bibitem{MendozaMunozElectro} 
Mendoza, M., Mu\~noz, J.D.
Three-dimensional lattice Boltzmann model for electrodynamics
(2010) Physical Review E - Statistical, Nonlinear, and Soft Matter Physics, 82 (5).
 \bibitem{ChopardDroz} Chopard, B., Droz M. 
 Cellular Automata Modeling of Physical Systems
 (1998) Cambridge University Press, Chapter 7. 
\bibitem{GuangwuWaves} 
Guangwu, Y.
A Lattice Boltzmann Equation for Waves
(2000) Journal of Computational Physics, 161 (1), pp. 61-69. 
 \bibitem{ref1}Favier J.,Revell A.,Pinelli A., \textit{A Lattice Boltzmann Immersed Boundary method to simulate the fluid interaction with moving and slender flexible objects}, J. of Comp. Phys., Volume 261, 145-161, 2014.
\bibitem{BGK}
Bhatnagar, P.L., Gross, E.P., Krook, M.
A model for collision processes in gases. I. Small amplitude processes in charged and neutral one-component systems
(1954) Physical Review, 94 (3), pp. 511-525. 
\bibitem{Elmore} Elmore W., Heald M. Physics of Waves, Dover Publications Inc., New York, 1969.
\bibitem{Kelvin}
J. Harvard.
The Design and Construction of Ships (1908): Vol. II: Stability, Resistance, Propulsion and Oscillations of Ships. P. 140.


\end{thebibliography}
\end{document}